\documentclass[floatfix,10pt,twocolumn,showpacs,preprintnumbers,amsmath,amssymb,aps,prb]{revtex4-1}
\pdfoutput=1

\usepackage{hyperref}
\usepackage{graphicx}
\usepackage{dcolumn}

\usepackage[usenames, dvipsnames]{color}

\def\mbf#1{\mathbf{#1}}
\def\c{c^\dagger}

\def\ket#1{|#1\rangle}

\begin{document}


\title{
The Dynamical Mean Field Theory phase space extension and\\
critical properties of the finite temperature Mott transition
}

\author{Hugo U. R. Strand}
 \email{hugo.strand@physics.gu.se}
\author{Andro Sabashvili}
\author{Mats Granath}
\author{Bo Hellsing}
\author{Stellan \"Ostlund}
\affiliation{%
University of Gothenburg, Gothenburg, Sweden
}%

\date{\today}

\begin{abstract}
 We consider the finite temperature metal-insulator transition in the half filled paramagnetic Hubbard model on the infinite dimensional Bethe lattice.
A new method for calculating the Dynamical Mean Field Theory fixpoint surface in the phase diagram is presented and shown to be free from the convergence problems of standard forward recursion. The fixpoint equation is then analyzed using dynamical systems methods.
On the fixpoint surface the eigenspectra of its Jacobian is used to characterize the hysteresis boundaries of the first order transition line and its second order critical end point.
The critical point is shown to be a cusp catastrophe in the parameter space, opening a pitchfork bifurcation along the first order transition line, while the hysteresis boundaries are shown to be saddle-node bifurcations of two merging fixpoints.
Using Landau theory the properties of the critical end point is determined and related to the critical eigenmode of the Jacobian.
Our findings provide new insights into basic properties of this intensively studied transition.
\end{abstract}

\pacs{71.30.+h, 71.10.Fd, 71.27.+a}
\maketitle

\section{\label{sec:introduction}Introduction}

The correlation driven Metal-Insulator Transition (MIT) at finite temperature, also known as the Mott transition, is today one of the most intensively studied phase transitions in solid state physics. The problem contains competing energy scales making it inaccessible to perturbative methods. The seminal work of Metzner and Vollhardt\cite{Metzner:1989aa} spurred a rapid development in this field by introducing the limit of infinite connectivity. In this limit the Dynamical Mean Field Theory (DMFT)\cite{Georges:1996aa, Kotliar:2006aa} becomes exact and the lattice problem can be mapped to an auxiliary impurity model connected to a non-interacting bath.

\begin{figure}
	\includegraphics{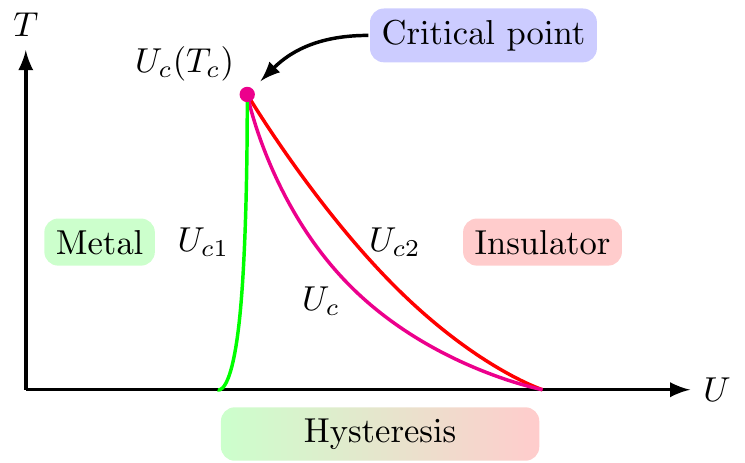}
	\caption{\label{fig:PhaseDiagSketch} (Color online)
	Sketch of the $(U,T)$ phase diagram showing 
        the first order thermodynamic transition line $U_{c}(T)$
        and the instability edges
        of the insulating and metallic solutions, $U_{c1}(T)$ and $U_{c2}(T)$ respectively.}
\end{figure}

The paramagnetic MIT of the Hubbard model on the Bethe lattice where spatial correlations and magnetic order parameters are suppressed, has been studied by many authors in this particular limit.\cite{Joo:2001rt, Bulla:2001yq, Capone:2007aa, Blumer:2002aa} Regarding this transition driven by Hubbard repulsion $U$, the emerging consensus is that it is a first-order phase transition terminating at a critical point. The low temperature first-order transition line $U_c(T)$ is surrounded by a hysteresis region with borders $U_{c1}(T)$ and $U_{c2}(T)$, containing both a metallic and an insulating solution. At a critical temperature $T_c$, the lines $U_{c1}(T)$, $U_{c2}(T)$ and $U_c(T)$ all meet in the second-order critical end point, $(U,T)= (U_c(T_c),T_c)$, as schematically shown in Fig.\ \ref{fig:PhaseDiagSketch}.

A theoretical framework for the understanding of the critical point has been presented by Kotliar {\it et.\ al.},\cite{Kotliar:1999aa,Kotliar:2000aa} explaining it in terms of a DMFT Landau functional and an emerging zeroth mode in the ``fluctuation matrix'' of this functional. Moreover, the double occupancy $D$ act as the thermodynamic conjugate variable to the (external) field $U$.\cite{Tong:2001aa}

In this paper we show, using the impurity solvers Exact Diagonalization (ED) and Iterated Perturbation Theory (IPT), that this theory also can explain the existence of a third thermodynamically unstable solution in the hysteresis region, previously reported by Tong and co-workers.\cite{Tong:2001aa}
We also present a general algorithm for finding fixpoints to the DMFT equations that is free from the stability and convergence problems encumbering both forward recursion\cite{Kotliar:2002aa,Blumer:2002aa} and Newton methods in the vicinity of the hysteresis boundaries and the critical point. The algorithm is quite general and can be implemented with any DMFT impurity solver.
Furthermore we present a method to calculate the Jacobian of the DMFT recursion relation at a fixpoint in the framework of ED and IPT. The properties of the Jacobian is then used to explain the origin of the numerical problems of forward recursion and Newton methods and how these difficulties are avoided by our algorithm.

This paper is organized as follows: In Section \ref{sec:Theory} we give an introduction to the single band Hubbard model on the Bethe lattice, in Section \ref{sec:DMFT} we introduce DMFT and how it can be reformulated as a fixpoint problem. In Section \ref{sec:ED} we present our implementation of the Exact Diagonalization impurity solver and how the Jacobian of the DMFT fixpoint function is calculated in this context. Section \ref{sec:IPT} is used to explain the same details for the Iterated Perturbation Theory impurity solver. Based on the general fixpoint problem we investigate the local convergence properties of the mentioned fixpoint solvers in Section \ref{sec:FixpointSolvers}. In Section \ref{sec:Thermodynamics} we give a brief description of the thermodynamics of the MIT. In Section \ref{sec:Results} the results are presented and put into relation with previous work in Section \ref{sec:Discussion}. Finally we give a short conclusion in Section \ref{sec:Conclusion}.

\section{\label{sec:Theory}Theory}
	
The Hamiltonian $\hat{H}$ of the half-filled Hubbard model is given by,
\begin{align}
	\hat{H} = &
	-t\sum_{<ij>,\sigma}
		\left( 
			c^\dagger_{i\sigma} c_{j\sigma} + 
			c^\dagger_{j\sigma} c_{i\sigma}
		\right) + \nonumber \\ &
                + U \sum_i c^\dagger_{i\uparrow} c_{i\uparrow}
                c^\dagger_{i\downarrow} c_{i\downarrow} 
                - \mu \sum_{i\sigma} c^\dagger_{i\sigma} c_{i\sigma}
                 \, ,
		\label{eq:HubbardModel}
\end{align}
with nearest neighbor hopping $-t$, local Hubbard repulsion $U$ and the chemical potential, $\mu = U/2$. In the limit of infinite dimensions, $d \rightarrow \infty$, the hopping matrix element $t$ has to be rescaled as, $t \rightarrow t/\sqrt{d}$, in order to retain a finite kinetic energy.\cite{Georges:1996aa}
On the Bethe lattice the non-interacting density of states $\rho^{(0)}(\omega)$ is semicircular and given by, 
\begin{equation}
	\rho^{(0)}(\omega) = \frac{2}{\pi} \sqrt{1 - \left(\frac{2\omega}{W}\right)^2}
	\, ,  \quad |\omega| < \frac{W}{2} \,,
\end{equation}
where $W$ is the bandwidth $(W = 4t)$. In this study we use, $W = 2\,$eV.
Since the Bethe lattice is bipartite the ground state of $\hat{H}$ is antiferromagnetic at low temperature, which should in principle suppress the MIT of the paramagnetic state studied here.  In the spirit of previous studies\cite{Georges:1996aa} we enforce the paramagnetic state by imposing translational invariance ignoring spatial correlations. This also enables us to connect to experimental results on more frustrated multi-band systems not displaying the antiferromagnetic instability.

\subsection{\label{sec:DMFT}Dynamical Mean Field Theory}

\begin{figure}
  \includegraphics{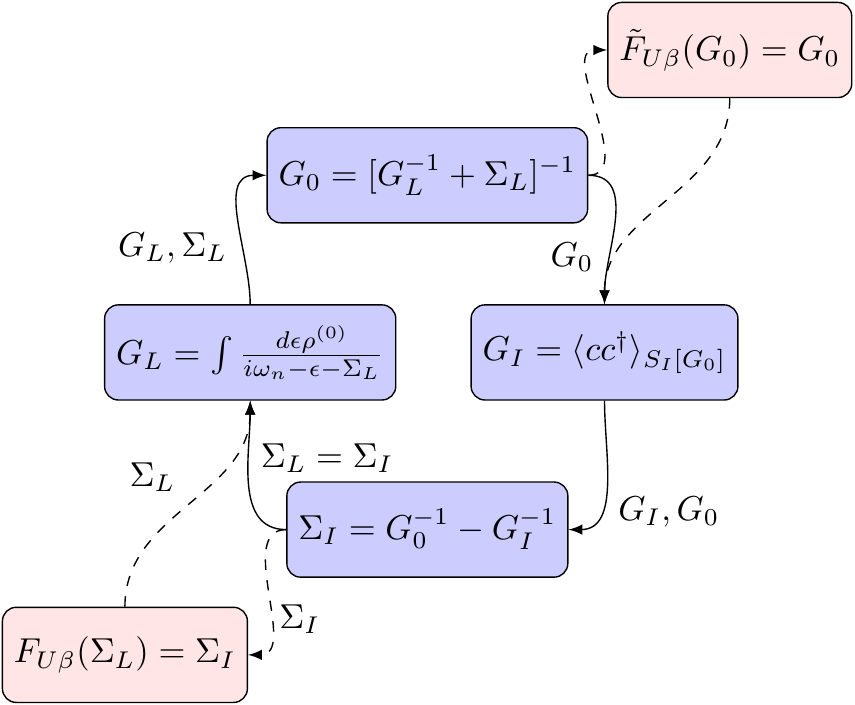}
  \caption{\label{fig:DMFTeq} (Color online)
    The self consistent DMFT equations and the two possible fixpoint function formulations $F_{U\beta}(\Sigma_L)$ and $\tilde{F}_{U\beta}(G_0)$.}
\end{figure}

Let us first introduce some concepts used in the formulation of DMFT. In the limit of infinite coordination the lattice self-energy $\Sigma_L$ is local, $\Sigma_L(\mbf{R}, i\omega_n) \rightarrow \Sigma_L(i\omega_n)$, and DMFT\cite{Georges:1996aa, Kotliar:2006aa} is an exact theory. In the continuum limit the local lattice Green's function, $G_L(\mbf{R} = \mbf{0}, i\omega_n) = G_L(i\omega_n)$, is given by,
\begin{equation}
	G_L ( i\omega_n ) =
	\int d\omega \,
	\frac{\rho^{(0)}(\omega)}
	{i\omega_n - \omega + \mu - \Sigma_L(i\omega_n)} 
        \, ,
	\label{eq:G_L}
\end{equation}
where $i\omega_n$ are Matsubara frequencies. With the lattice we associate an auxiliary impurity connected to a bath acting as a dynamic Weiss field. The bath Green's function $G_0$ is obtained by subtracting the local interactions using $\Sigma_L$.
\begin{equation}
	G_0(i\omega_n) = \left[ G_L^{-1}(i\omega_n) + \Sigma_L(i\omega_n) \right]^{-1}
	\label{eq:G_0}
\end{equation}
The local interactions of the lattice Hamiltonian $\hat{H}$ and $G_0$ now fully determine the  action $S_I$ for the impurity system,
\begin{eqnarray}
	S_I[G_0] = 
	U \int_0^\beta d\tau \, 
	\c_\uparrow(\tau) c_\uparrow(\tau) 
	\c_\downarrow(\tau) c_\downarrow(\tau)
	- \nonumber \\ 
	\quad- \int_0^\beta d\tau \int_0^\beta d\tau' 
	\sum_\sigma \c_\sigma(\tau) G_0^{-1}(\tau - \tau') c_\sigma(\tau')  \, .
	\label{eq:ActionS_I}
\end{eqnarray}
Solving the impurity problem is still a formidable task but the single impurity system is now well within reach for state of the art numerical algorithms. Using an impurity solver the paramagnetic interacting impurity Green's function, $G_I(i\omega_n)$, can be calculated as,
\begin{align}
	G_{I\sigma \sigma'}(i\omega_n) = & - \int_0^\beta d\tau\, e^{i\omega_n \tau}
        \langle T c_\sigma(\tau) c_{\sigma'}^\dagger(0)
        \rangle_{S_I[G_0]}
        \nonumber \, ,\\
       \textrm{where, } 
       G_{I} =  & \, G_{I\uparrow\uparrow} = G_{I\downarrow\downarrow} 
       \,, \quad G_{I\uparrow\downarrow} = G_{I\downarrow\uparrow}
       = 0 
       \, .
	\label{eq:G_I}
\end{align}
The corresponding impurity self-energy $\Sigma_I$ is calculated by inverting the Dyson equation,
\begin{equation}
	\Sigma_I(i\omega_n) = G_0^{-1}(i\omega_n) - G_I^{-1}(i\omega_n)
	\label{eq:S_I}
        \, .
\end{equation}
Now given $\Sigma_L$, the equations, (\ref{eq:G_L}, \ref{eq:G_0}, \ref{eq:G_I} and \ref{eq:S_I}), can be used to calculate a corresponding $\Sigma_I$. Incorporating these steps into a single DMFT function $F$ gives the short form,
\begin{equation}
	F_{U\beta}(\Sigma_L) = \Sigma_I
        \, ,
\end{equation}
where the Hubbard $U$ and the inverse temperature, $\beta = 1/k_BT$, are external parameters.

The last step is to find self consistent solutions where the lattice and impurity self-energies coincide, $\Sigma_L(i\omega_n) = \Sigma_I(i\omega_n)$. This is equivalent to finding fixpoint solutions $\Sigma^*$ of the DMFT function $F_{U\beta}$,
\begin{equation}
	F_{U\beta}(\Sigma^*) = \Sigma^* \, , \quad \Sigma^* = \Sigma_L = \Sigma_I \, .
	\label{eq:FixedPoint}
\end{equation}
In principle any fixpoint algorithm for multidimensional functions can now be applied on $F_{U\beta}$ to find DMFT solutions. 
The choice of $\Sigma$ as the fundamental variable in the fixpoint scheme is not unique, as one can alternatively use the bath Green's function $G_0$ as fixpoint variable giving fixpoint solutions, $\tilde{F}_{U\beta}(G_0^*)=G_0^*$. Both possibilities are indicated in Fig.\ \ref{fig:DMFTeq}, where the coupled equations forming the DMFT fixpoint functions are shown schematically.

\subsection{\label{sec:ED}Exact Diagonalization}

\begin{figure}
  \includegraphics{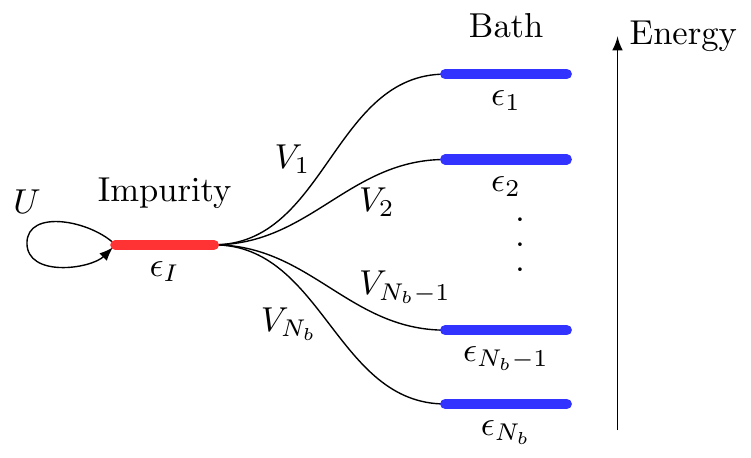}
  \caption{\label{fig:AndMod} (Color online)
    Truncated SIAM, with one correlated impurity level $\epsilon_I$, bath levels $\epsilon_k$, $1\le k \le N_b$, hybridizations $V_k$ and Hubbard repulsion $U$.}
\end{figure}

To solve the impurity problem of Eqs.\ (\ref{eq:ActionS_I}) and (\ref{eq:G_I}) we have implemented
\footnote{The Exact Diagonalization code has been developed by the
  corresponding author using Python,\cite{Rossum:1995aa} and the
  numerical modules Numpy and Scipy,\cite{Jones:2001aa} providing both
  conjugate gradient minimizers and Broyden's method. The
  diagonalization was performed using LAPACK and the MPI
  parallelization over spin sectors was implemented using the module mpi4py. Finally the Lehmann representation sum, eq. (\ref{eq:Lehmann}), was implemented in FORTRAN and interfaced with Python using f2py.\cite{Peterson:2009aa}} 
the Exact Diagonalization (ED) algorithm by Caffarel and Krauth.\cite{Caffarel:1994aa}
In ED the impurity problem is projected to a truncated Single Impurity Anderson Model (SIAM), with the Hamiltonian,
\begin{align}
	\hat{H}_{\textrm{\tiny SIAM}} = &
	\sum_\sigma (\epsilon_I - \mu) c^\dagger_{\sigma} c_{\sigma} + 
	\sum_{k\sigma} \epsilon_k c^\dagger_{k\sigma} c_{k\sigma} + 
	\nonumber \\ &
	\sum_{k\sigma} V_{k} 
	\left( c^\dagger_{\sigma} c_{k\sigma} + c^\dagger_{k\sigma}
          c_{\sigma}\right) + 
	U c^\dagger_{\uparrow} c_{\uparrow} c^\dagger_{\downarrow} c_{\downarrow}
        \, ,
\end{align}
keeping the local interaction of the lattice Hamiltonian $\hat{H}$ but replacing the hopping term with hybridization $V_k$ to a set of ``bath'' states at energies $\epsilon_k$, see Fig.\ \ref{fig:AndMod}.

As $\hat{H}$ is particle-hole symmetric the same symmetry is imposed on the SIAM parameters. Letting $\epsilon_I=0$, the bath level energies $\epsilon_k$ are placed symmetrically around zero energy. Moreover for each pair, $\epsilon_k = -\epsilon_{\tilde{k}} \ne 0$, the hybridizations are the same, $V_k = V_{\tilde{k}}$. Half filling is obtained by fixing the chemical potential $\mu$ to, $\mu = U/2$.

The first step of the algorithm is to project the impurity bath Green's function $G_0$ to the non-interacting ($U=0$) SIAM Green's function $G^{\textrm{\tiny SIAM}}_{0}$,
\begin{equation}
	G^{\textrm{\tiny SIAM}}_{0} [\epsilon_k,V_k](i\omega_n)
	=
	\left[
		i\omega_n + \mu - \epsilon_I
		- \sum_k \frac{|V_k|^2}{i\omega_n - \epsilon_k}
	\right]^{-1}
        \, ,
\end{equation}
through minimization of a penalty function $\chi^2$,
\begin{equation}
	\chi^2 [\epsilon_k,V_k] = \frac{1}{N} \sum_n^N 
	\left| 
		G_0(i\omega_n) - G_0^{\textrm{\tiny SIAM}}[\epsilon_k,V_k] (i\omega_n)
	\right|^2
	\, ,
        \label{eq:chi2}
\end{equation}
with respect to the SIAM parameters $\epsilon_k$ and $V_k$. To minimize $\chi^2$ a standard conjugate gradient minimization algorithm is used. Many forms of the penalty function $\chi^2$ can be constructed\cite{Koch:2008aa} and in this work we choose Eq.\ (\ref{eq:chi2}), which is sensitive to the low frequency behavior of $G^{\textrm{\tiny SIAM}}_{0}$, and $N=2^9$ matsubara frequencies.

With the parameters of $\hat{H}_{\textrm{\tiny SIAM}}$ determined its matrix representation in the occupation number basis is calculated. The symmetries of $\hat{H}_{\textrm{\tiny SIAM}}$ can be used to block-diagonalize the matrix representation. For simplicity we only exploit the symmetry that subspaces with fixed number of spin up $n_\uparrow$ and spin down $n_\downarrow$ are not mixed by $\hat{H}_{\textrm{\tiny SIAM}}$.
Let us denote the eigenstates of $\hat{H}_{\textrm{\tiny SIAM}}$ by $\ket{\nu}$ where, $\hat{H}_{\textrm{\tiny SIAM}} \ket{\nu} = E_\nu \ket{\nu}$.

The eigenstates are explicitly calculated by diagonalization and used to calculate the impurity Green's function $G_{I\sigma\sigma'}$ from the Lehmann spectral representation,\cite{Fetter:2003aa}
\begin{equation}
	G_{I\sigma\sigma'}(i\omega_n) = 
	\frac{1}{Z}
	\sum_{\nu,\mu}
	\frac{\langle \mu | c^\dagger_{\sigma} | \nu \rangle \langle \nu | c_{\sigma'} | \mu \rangle}
		 {i\omega_n + E_\mu - E_\nu}
	\left( e^{-\beta E_\nu} + e^{-\beta E_\mu} \right)
        \, .
        \label{eq:Lehmann}
\end{equation}
As the the system is paramagnetic, Eq.\ (\ref{eq:G_I}) holds, and only one impurity Green's function $G_I$ has to be calculated. The double occupancy $D$ is given by, 
\begin{align}
	D = & 
	\frac{1}{Z}
	\sum_{\nu}
	\langle \nu | \hat{n}_{\uparrow} \hat{n}_{\downarrow}| \nu \rangle 
	e^{-\beta E_\nu} 
        \, .
\end{align}

\begin{figure}
	\includegraphics{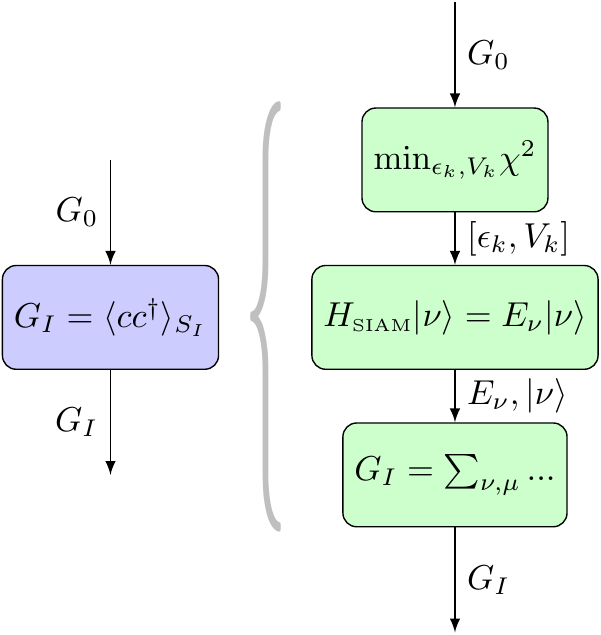}
	\caption{\label{fig:EDeq} (Color online)
	Detailed schematic (right) of the Exact Diagonalization impurity
        solver (left).}
\end{figure}

The described steps of the ED algorithm are schematically shown in Fig.\ \ref{fig:EDeq} and replaces the, $G_I = \langle c c^\dagger \rangle_{S_I[G_0]}$, block in the DMFT equations of Fig.\ \ref{fig:DMFTeq}.

In our calculations the dimension of the SIAM Hilbert space is the limiting factor of the ED algorithm. With one impurity level the number of fermionic states $N_f$ is given by, $N_f = N_b + 1$, where $N_b$ is the number of bath levels. The corresponding size of the Hilbert space becomes $2^{2N_f}$, growing exponentially with respect to $N_f$. Dividing the Hamiltonian in blocks of constant $(n_{\uparrow}, n_{\downarrow})$ gives a set of smaller Hilbert spaces with dimensions, 
$\binom{N_f}{n_\uparrow} \binom{N_f}{n_\downarrow}$.
Our calculations converge rapidly with the number of fermionic levels
$N_f$ and for, $N_f=6$, used in our calculations, the region around
the critical point is well converged. \cite{Liebsch:2005aa,Pozgajcic:2004lr}

\subsubsection{\label{sec:EDFixpoint}Fixpoint function and Jacobian}

We now discuss the coupled DMFT equations as a fixpoint function in terms
of $G_0$ instead of $\Sigma_L$. In the ED algorithm $G_0$ is parametrized by a small number of parameters, $\mbf{x} = [\epsilon_k, V_k]$, when projected on to the SIAM and a solution of the DMFT equations can be formulated as a fixpoint problem in $\mbf{x}$, $F_{U\beta}(\mbf{x}^*) = \mbf{x}^*$.

The reduction of parameters using $\mbf{x}$ instead of $G_0(i\omega_n)$ facilitates a direct calculation of the Jacobian, $J_F(\mbf{x}) = \nabla F_{U\beta}(\mbf{x})$. At each obtained fixpoint
$\mbf{x}^*$, $F_{U\beta}(\mbf{x}^*) = \mbf{x}^*$, the low dimensional parameter space allows us to use a modified central finite differences formula to calculate the Jacobian, 
\begin{align}
  J_F(\mbf{x}) \approx & 
 \frac{ F_{U\beta}(\mbf{x} +
    h \hat{\mbf{x}}_n) - F_{U\beta}(\mbf{x} -
    h \hat{\mbf{x}}_n) }{2 h}
\, ,
\end{align}
where $h$ is the discretization, $\hat{\mbf{x}}_n$ is the unit vector in the n:th dimension.
Due to the large parameter spread in $\mbf{x}^*$ relative scaling was applied to stabilize the numerical evaluation of $J_F(\mbf{x}^*)$.

\subsection{\label{sec:IPT}Iterated Perturbation Theory}

To show the generality of the fixpoint analysis we also consider the Iterated Perturbation Theory (IPT)\cite{Georges:1996aa} formulation of the DMFT equations, which amounts to solving the impurity problem, Eq.\ (\ref{eq:G_I}), perturbatively to second order in $U$. 
To discretize the problem in this case we consider directly the iteration scheme of the self-energy $\Sigma(\tau)$ defined on a discrete time mesh, $\tau_j=\frac{\beta}{N}j$ with $N$ a constant integer and $j$ integer. Clearly the step size $\beta/N$ needs to be increased with decreasing temperature to be able to capture the self-energy or Green's function sufficiently well. Because of the discretization in time, we can represent the self-energy $\Sigma(\tau_j)$ (or Green's function) using a finite number of Matsubara frequencies, $\omega_n=\frac{2\pi}{\beta}(n+\frac{1}{2})$, with, $0\leq n<N$. 
Thus defining a ``Matsubara-periodized'' self-energy $\Sigma(i\omega_n)$ through, 
$\Sigma(\tau_j)={\beta}^{-1}\sum_{n=0}^{N-1}e^{-i\omega_n\tau_j}\Sigma(i\omega_n)$, and,
$\Sigma(i\omega_n)=\frac{\beta}{N}\sum_{j=0}^{N-1}e^{i\omega_n\tau_j}\Sigma(\tau_j)$, 
which is now periodic with period $\frac{2\pi}{\beta}N$. We formulate a fixpoint equation in terms of the finite dimensional self-energy similarly, 
\begin{equation}
  F_{U\beta}\left(\Sigma(i\omega_n)\right) = \Sigma'(i\omega_n)
  \, ,
\end{equation}
allowing for the same study of the fluctuations around the fixpoint as for the ED formulation but now in the $N$ dimensional space spanned by $\Sigma(i\omega_n)$. Specifically we calculate the Jacobian of $F_{U\beta}$ through,
\begin{align}
  J_F(\Sigma) 
\approx &
  \frac{F_{U\beta}(\Sigma + h\hat{z}_n) - F_{U\beta}(\Sigma - h\hat{z}_n)}{2h}
  \, , 
\end{align} 
where, $\hat{z}_n = i(\hat{x}_n - \hat{x}_{-n-1})/\sqrt{2}$, is a unit vector in the particle-hole symmetric $\Sigma(i\omega_n)$ subspace and $h$ is the finite difference discretization.

We will not present the details of the calculations here in terms of ``Matsubara-periodized'' Greens functions\footnote{A. Sabashvili {\it et.\ al.\ } in preparation} but only point out some main features. Since we have discretized the Green's function, we have to be particular careful about defining, $G(\tau_j = 0)$. Particle-hole symmetry implies that, $G_L(\tau) = - G_L(-\tau)$, and to preserve this we define, $G(\tau_j = 0 ) = 0$. With this definition the first-order Hartree-contribution to $\Sigma$ is zero and correspondingly, $\mu = 0$, at half-filling.
The discretized IPT approximation for the self-energy is given by,  $\Sigma(\tau_j)=-U^2G^2(\tau_j)G(-\tau_j)=U^2G^3(\tau_j)$, with $G(\tau_j=0)=0$ by definition. From this follows also that $\Sigma$ is purely imaginary.
 
The DMFT equations can be written exactly in terms of the periodized Green's function and self-energy by replacing Eq.\ (\ref{eq:G_L}) with,
\begin{equation}
G_L(i\omega_n)=\int
d\omega\frac{\rho^0(\omega)}{\frac{2N}{\beta}(\coth
  \frac{\beta}{2N}(i\omega_n-\omega))^{-1}-\Sigma(i\omega_n)}
\, ,
\end{equation}
where the $\coth(...)$ term is the exact expression for the discrete Fourier transform of the non-interacting Green's function on the lattice.  

\subsection{\label{sec:FixpointSolvers}Fixpoint solvers}

A common ingredient in all DMFT calculations is solving a fixpoint problem. Conceptually the choice to parametrize the fixpoint function in terms of $\Sigma$ or $G_0$ is irrelevant, since the resulting fixpoints are equivalent whichever coordinates are used. In this section we adapt $\Sigma$ as the fixpoint-variable keeping in mind that it is exchangeable with $G_0$ (or $\mbf{x}=[\epsilon_k, V_k]$).

We now discuss the two most widely used algorithms for solving the fixpoint problem, forward recursion and Newton methods and for each method the local convergence properties around a
fixpoint $\Sigma^*$ will be explained in terms of the dominating eigenvalue $\epsilon$ of the Jacobian, $J_F(\Sigma^*)$. Finally we introduce the phase space extension and explain why this method is free from some of the deficiencies of forward recursion and Newton methods.

\subsubsection{Forward recursion}

The most common algorithm for solving the DMFT equations is the \emph{fixpoint forward recursion}.\cite{Georges:1996aa} Given an initial guess $\Sigma_0$ a series $\{\Sigma_n\}$ is generated by the recursion relation,
\begin{equation}
	\Sigma_{n+1} = F_{U\beta}(\Sigma_{n})
        \, ,
\end{equation}
and a fixpoint $\Sigma^*$ is found if the series converges, $\Sigma^* = \Sigma_\infty$. To study the convergence properties of the series $\{ \Sigma_n \}$ in the vicinity of a fixpoint $\Sigma^*$, we can approximate $F_{U\beta}$ by its first order Taylor expansion,
\begin{equation}
	F_{U\beta}(\Sigma^* + \delta \Sigma) \approx F_{U\beta}(\Sigma^*) + J_F(\Sigma^*) \cdot \delta\Sigma
        \, ,
\end{equation}
where $J_F$ is the Jacobian matrix of $F_{U\beta}$, $J_F(\Sigma) = \nabla F_{U\beta}(\Sigma)$ and $\delta \Sigma$ is a small perturbation. By repeated application of the recursion near the fixpoint,
\begin{align}
	\Sigma_0  = & \Sigma^* + \delta \Sigma \nonumber \\
	\Sigma_n = & \Sigma^* +  J_F(\Sigma^*)^n \delta \Sigma 
        \, ,
\end{align}
we easily observe that the convergence is determined by the eigenvalues of $J_F$ and in particular by the eigenvalue, $\epsilon$, with the largest magnitude. Hence we require, $|\epsilon|<1$, for forward recursion to converge at all. This imposes a restriction on the solution space that can be found by this scheme.

If the Jacobian at a fixpoint has an eigenvalue larger than one in magnitude, the algorithm will only converge if the perturbation $\delta\Sigma$ has no components in the corresponding eigenspace. Any contribution in $\delta \Sigma$ from this eigenspace will be amplified and $\Sigma_n$ will move away from the fixpoint $\Sigma^*$, and the forward recursion algorithm will be unable to find the fixpoint in the first place. Fixpoint forward recursion can therefore only be used to find a subset of all fixpoints $\Sigma^*$ of the function $F_{U\beta}$ whose Jacobian have all eigenvalues bounded by one in magnitude.
And if, $|\epsilon| \rightarrow 1^-$, when tuning an external parameter the fixpoint forward recursion experiences a \emph{critical slowing down} of convergence, due to the damping factor $\epsilon^n$ of a perturbation. This phenomena has been reported for the MIT of DMFT when approaching the hysteresis boundaries of the phase diagram.\cite{Kotliar:2002aa,Blumer:2002aa} 

\subsubsection{Newton methods}

The family of Newton's method and the quasi Newton methods\cite{Heath:2002aa} are all multi dimensional root solvers with better stability properties than forward recursion. Broyden's method\cite{Broyden:1965aa} from this class of algorithms have recently been applied in the context of DMFT.\cite{Zitko:2009aa} In order to use a root solver, the fixpoint problem in Eq.\ (\ref{eq:FixedPoint}) is simply reformulated to a root problem, 
\begin{equation}
	R_{U\beta}(\Sigma^*) \equiv F_{U\beta}(\Sigma^*) - \Sigma^* = \mbf{0}
        \, ,
        \label{eq:rootproblem}
\end{equation}
where the Jacobian $J_R$ of $R_{U\beta}$ has the form,
\begin{equation}
	J_R(\Sigma) = \nabla R_{U\beta}(\Sigma) = \nabla F_{U\beta}(\Sigma) - \mbf{1} 
	= J_F(\Sigma) - \mbf{1}
        \, .
\end{equation}

The series $\{\Sigma_n\}$ is in the case of Newton's method generated as,
\begin{equation}
  \Sigma_{n+1} = \Sigma_n - (J_R(\Sigma_n))^{-1} R_{U\beta}(\Sigma_n)
  \, .
\end{equation}

In the linear regime in the vicinity of a fixpoint $\Sigma^*$ where,
$R_{U\beta}(\Sigma^*+\delta \Sigma) \approx J_R(\Sigma^*) \delta \Sigma$ and
$J_R(\Sigma^*+\delta \Sigma) \approx J_R(\Sigma^*)$, the series converges in one iteration as,
\begin{align}
	\Sigma_0 =  & \Sigma^* + \delta \Sigma \nonumber \\
	\Sigma_1 = & \Sigma_0 - J_R^{-1}(\Sigma_0) R_{U\beta}(\Sigma_0) \approx
        \Sigma^*
        \, ,
\end{align}
assuming that $J_R(\Sigma^*)$ is invertible. Translated to the eigenvalue spectrum of the Jacobian $J_F(\Sigma^*)$ of the fixpoint function $F_{U\beta}(\Sigma^*)$, local convergence is achieved as long as no eigenvalue is equal to one, a less restrictive requirement compared to forward recursion. Newton's method will therefore converge even in areas of parameter space where forward recursion fails completely.

\subsubsection{The phase space extension}

Although Newton's metod allows us to work with, $|\epsilon|>1$, a
problem remains when, $|\epsilon| = 1$, which (as we will show) is precisely at the
hysteresis boundaries of the MIT.
To be able to trace the solutions across this singularity we reformulate the problem
in an extended phase space where the resulting Jacobian no longer becomes singular at the hysteresis boundaries. 

We construct a real-valued function $A(\Sigma)$ whose value lifts the degeneracy of the coexisting fixpoints of the DMFT fixpoint function. Using $A(\Sigma)$ we write down an extended root problem that not only finds a DMFT solution but also fixes the value of $A$ to some given value $\alpha$, $A(\Sigma) = \alpha$.
By increasing the dimension of the root problem by one, the extended root function $\tilde{R}_{\alpha\beta}$ can be defined as,
\begin{equation}
	\tilde{R}_{\alpha \beta}(U, \Sigma) = 
	( \alpha - A(\Sigma) , R_{U\beta}(\Sigma) ) =
        \mbf{0}
        \, ,
\end{equation}
where $U$ is treated as a free parameter to be varied along with $\Sigma$ in order to obtain, $ \alpha - A(\Sigma) = 0$. By constraining the value of $A(\Sigma)$ to $\alpha$ the degeneracy of the fixpoints is lifted and the resulting Jacobian $J_{\tilde{R}}$ is invertible even at the hysteresis boundaries.

In this investigation, $A(\Sigma) = \textrm{Im}[\Sigma(i\omega_0)]$, was used and sampled on a logarithmic grid. The extended root problem was solved using Broyden's second method.\cite{Broyden:1965aa} As Broyden's method does not require any Jacobian evaluations it has the same computational cost as forward recursion, but with an almost constant convergence rate in the entire phase diagram.

\subsection{\label{sec:Thermodynamics}Thermodynamics}

Studying the MIT in the $(U,T)$ phase space requires understanding of
the thermodynamics of $\hat{H}$ in Eq.\ (\ref{eq:HubbardModel}). Let us first consider the expectation value $\langle \hat{H} - \mu \hat{N} \rangle$ and introduce some notation,
\begin{align}
	\langle \hat{H} - \mu \hat{N} \rangle = &
	\langle \hat{T} \rangle + U\sum_i \langle \hat{D}_i \rangle 
        - \mu \sum_{i\sigma} \langle \hat{n}_{i\sigma} \rangle 
        = \\ = &
	\langle \hat{T} \rangle + 
        UN \left( \langle D \rangle - \frac{1}{2} \right)
        \, ,
\end{align}
where $N$ is the number of sites, $\hat{D}_i = \hat{n}_{i\uparrow} \hat{n}_{i\downarrow}$ is the on site double occupancy, and the single particle hopping is contained
in the kinetic term $\hat{T}$. In the last step we have assumed half-filling with $\langle \hat{n}_{i\sigma} \rangle = 1/2$ and $\mu = U/2$. It is now evident that $U$ acts as an external field and can be considered a conjugate variable to the double occupancy, $D = \langle \hat{D} \rangle$. From this it is possible to derive a Maxwell construction in $U$ and $D$ analogous to the formulation in density and pressure for the van der Waals equation of state.\cite{Reichl:2004aa, Tong:2001aa}

From the definition of the grand partition function $\mathcal{Z}(\beta, \mu, U)$ and the free energy $\Omega(\beta, \mu, U)$,
\begin{eqnarray}
  \mathcal{Z} = e^{-\beta \Omega} = \textrm{Tr} \left[
    e^{-\beta(\hat{H} - \mu \hat{N})} \right]
  \, ,
\end{eqnarray}
the derivative of $\Omega$ with respect to $U$ is given by,
\begin{eqnarray}
  \left.\frac{\partial \Omega}{\partial U}\right|_{\beta} = 
  -\frac{1}{\beta} \frac{\partial}{\partial U} \ln \mathcal{Z}
 =  N \left( \langle \hat{D} \rangle - \frac{1}{2} \right)
  \, .
\end{eqnarray}
The free energy difference $\Delta\Omega$ between two points on an isotherm can be expressed as,
\begin{equation}
  \frac{\Delta \Omega}{N} = \int_{U_1}^{U_2} dU\, 
 \left( \langle \hat{D} \rangle - \frac{1}{2} \right)
  \, .
  \label{eq:Maxwell}
\end{equation}
provided that there is an adiabatic connection between $U_1$ and $U_2$. 

In the case of the MIT this can be used to determine the thermodynamic first order transition given by the three DMFT solutions on an isotherm in the hysteresis region, as reported previously by Tong {\it et.\ al. }\cite{Tong:2001aa} This third unstable solution connects the metallic and insulating solutions making it possible to calculate the free energy difference between the metal and insulator for a given Hubbard U.

\section{\label{sec:Results}Results}

\begin{figure}
  \includegraphics{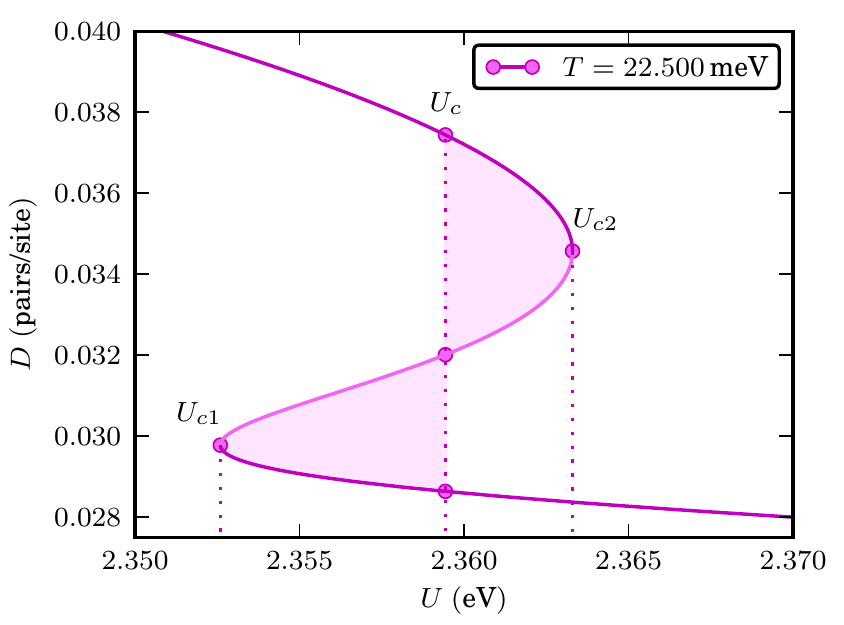}
  \caption{\label{fig:Maxwell} (Color online)
    Isotherm at, $T=22.50\,$meV, inside the hysteresis region, with the double occupancy $D$ as a function of $U$. Shaded areas show the Maxwell construction. The solid line is composed of dense DMFT-ED solution points.}
\end{figure}

Using the phase space extension and Broyden's second method to solve the DMFT equations we find three solutions in the hysteresis region at fixed $U$ and $\beta$. One metallic solution with high double occupancy $D$, one insulating solution with low $D$ and a third intermediate ``unstable'' solution, see Fig.\ \ref{fig:Maxwell}. As a function of $U$ these three solutions form a continuous Z-shaped isotherm in $D(U)$, where the unstable solution adiabatically connects the metallic and insulating solutions.
The unstable solution is not an artifact due to the phase space extension since it also is a solution to the DMFT root problem, $R_{U\beta}(\Sigma^*) = \mbf{0}$, of equation (\ref{eq:rootproblem}).

With the continuous isotherm $D(U)$, of Fig.\ \ref{fig:Maxwell}, it is possible to apply the Maxwell construction, Eq.\ (\ref{eq:Maxwell}), to determine the thermodynamic first order transition $U_c(T)$. Where the free energies of the metallic and insulating solutions coincide.
This corresponds to equating the enclosed areas left and right of $U_c$, as shown in Fig.\ \ref{fig:Maxwell}. It is evident that the unstable solution always has a free energy higher than both the metallic and insulating solution, thus always being thermodynamically unstable. Increasing the temperature towards the critical temperature $T \rightarrow T_c^-$ shrinks the size of the hysteresis region and at $T=T_c$ it disappear.

Many of our results can now bee seen to be quite general properties of the MIT. The DMFT fixpoint solutions form a continuous surface in the $(U,T,D)$ phase space. At the critical point $(U_c,T_c,D_c)$ this surface has a cusp singularity. As a function of the DMFT recursion this surface opens up a pitchfork bifurcation\cite{Crawford:1991qy} in the $(T,D)$ plane and two of the solutions annihilate by saddle-node bifurcations\cite{Crawford:1991qy} at the hysteresis boundaries $U_{c1}(T)$ and $U_{c2}(T)$ in the $(U,T)$ plane.
%
\begin{figure}
  \includegraphics{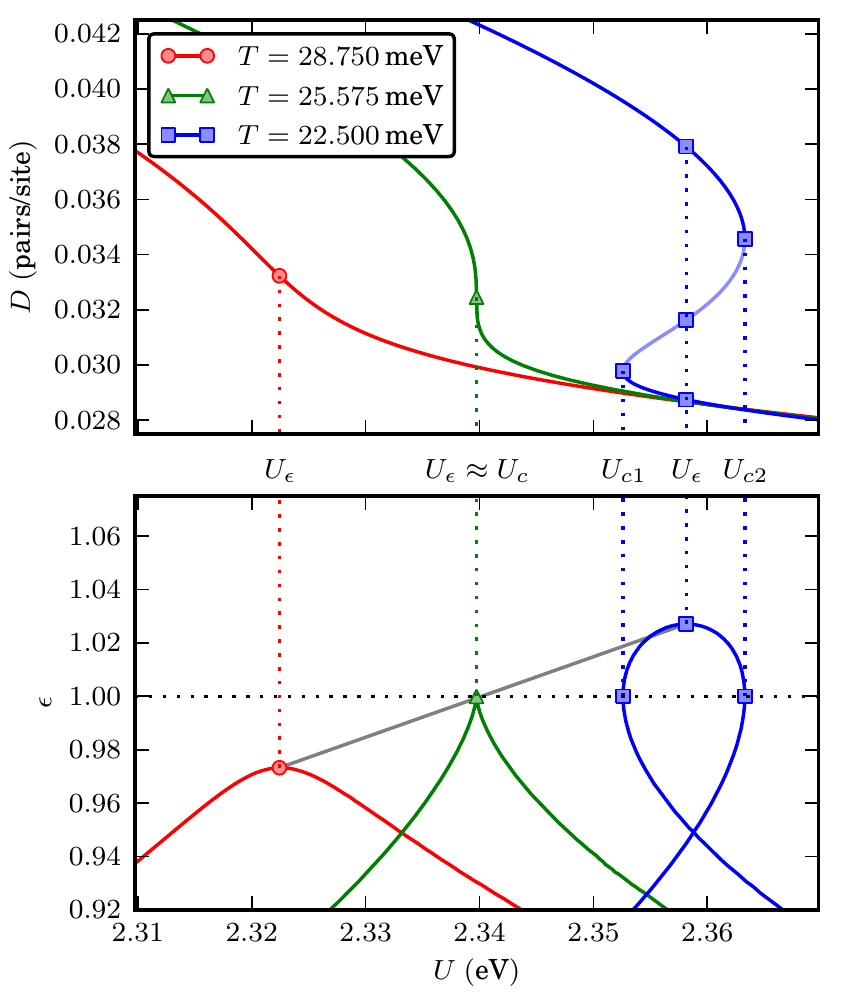}
  \caption{\label{fig:DEvVSU} (Color online)
    Double occupancy $D$ (upper panel) and maximum eigenvalue $\epsilon$ of $J_F(\Sigma^*)$ (lower panel) plotted against $U$ on isotherms above, close to and below the critical point (\textcolor{red}{circles}, \textcolor{OliveGreen}{triangles} and \textcolor{blue}{squares} respectively). Saddle-node bifurcation boundaries $U_{c1}$ and $U_{c2}$ occur when $\epsilon = 1$. The solid lines are composed of dense DMFT-ED solution points.}
\end{figure}
%
Studying the isotherms of $D(U)$ and $\epsilon(U)$, where $\epsilon$ is the in magnitude largest eigenvalue of $J_F(\mbf{x}^*)$, around the critical point, see Fig.\ \ref{fig:DEvVSU}, we can
explain the behavior of the common algorithms used to solve the DMFT equations.

Above the critical temperature, $T>T_c$, the eigenvalue $\epsilon$ is always less than one and both forward recursion and Newton's method converge. In this regime the maximum of $\epsilon$ at the coupling $U_\epsilon$ determines the center of the thermodynamic crossover region.\cite{Rozenberg:1995zr,Bulla:2001yq}
At the critical temperature, $T\sim T_c$, forward recursion displays a \emph{critical slowing down} of convergence, as $\epsilon \rightarrow 1^-$ when $U \rightarrow U_c$, while Newton's method becomes unstable first at the critical point $(U_c,T_c)$. 
Below the critical temperature, $T < T_c$, the behavior of $\epsilon(U)$  becomes more complicated. Following the metallic solution (high $D$) the solution annihilate with the unstable solution at the second hysteresis boundary $U_{c2}$ through a saddle-node bifurcation. This coincides with $\epsilon \rightarrow 1^-$ and explains the critical slowing down of forward recursion when approaching the hysteresis boundary from the inside of the hysteresis region. The behavior of the insulating solution (low $D$) at the first hysteresis boundary $U_{c1}$ is analogous to that of the metallic solution.

The unstable solution emerges at the hysteresis boundaries $U_{c1}$ and $U_{c2}$ through the saddle-node bifurcations and has $\epsilon \ge 1$ in the entire hysteresis region. Thus forward recursion will never find this solution, although we have been able to trace it using Newton's method by supplying a close enough initial guess. But Newton's method is still unstable on the hysteresis boundaries and a far better method is the phase space extension which converges everywhere in the studied parameter range.

Recalling the definition of $\epsilon(U,T)$ as the maximum eigenvalue of $J_F(\Sigma^*)$ let us define $U_\epsilon(T)$ to be the coupling that maximizes $\epsilon$ for fixed $T$, see Fig.\ \ref{fig:DEvVSU}.
This allows for a precise determination of the critical temperature $T_c$ and critical coupling $U_c(T_c)$ as, $\epsilon(U_\epsilon(T_c), T_c) = 1$ and $U_c(T_c) = U_\epsilon(T_c)$. By linear interpolation of $\epsilon(U_\epsilon(T), T)$ from isotherms above and below $T_c$ the critical end point can be directly determined, as indicated in the lower panel of Fig.\ \ref{fig:DEvVSU}. Away from $T_c$, $U_\epsilon(T)$ is not equal to the coupling $U_c(T)$ where the first order transition occur.

\begin{figure}
  \includegraphics{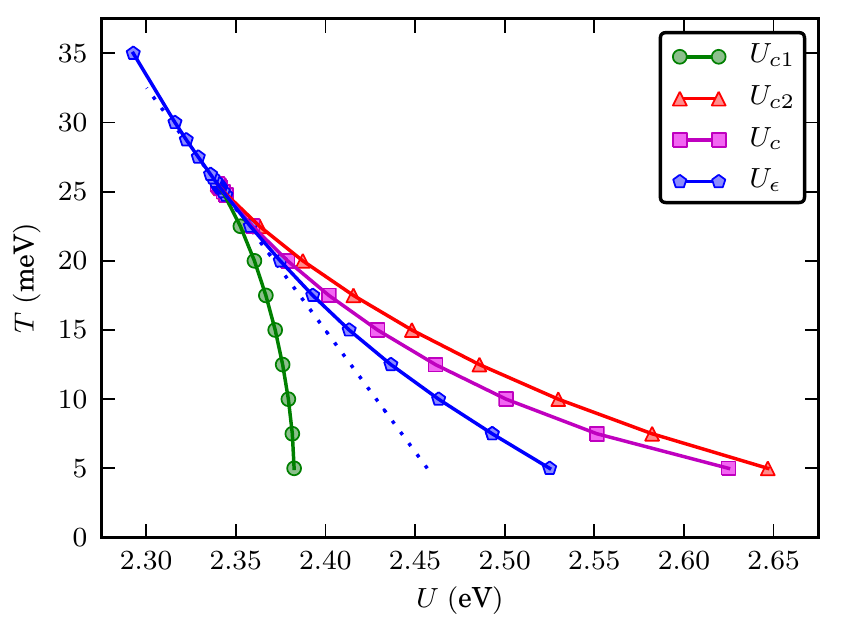}
  \caption{\label{fig:PhaseDiag} (Color online)
    Phase diagram in $U$ and $T$ plane, markers are ED-DMFT data and the dotted line correspond to, $a_0(u,t) = 0$, to linear order i.e., $\beta_0 u + \gamma_0 t =0$. (Solid lines are guides for the eye.)}
\end{figure}

The $(U,T)$ phase diagram can now be represented by four lines, see Fig.\ \ref{fig:PhaseDiag}, the hysteresis boundaries $U_{c1}(T)$ and $U_{c2}(T)$ that confine the region of fixpoint coexistence, the first order thermodynamic transition line $U_c(T)$  between metal and insulator and the maximal eigenmode curve $U_\epsilon(T)$. Where $U_\epsilon(T)$  gives the crossover between metal and insulator for temperatures above the critical temperature $T_c$.

To accurately determine the location $(U_c(T_c),T_c,D_c)$ of the  critical end point and its critical properties we fit a Landau functional model $\mathcal{L}$ to the calculated DMFT-ED isotherms near the critical end point,
\begin{equation}
  \mathcal{L}(u,t,d) = 
  a_0 d + a_1 \frac{d^2}{2} + a_2 \frac{d^3}{3} + \frac{d^4}{4}
  \label{eq:Landau}
  \, ,
\end{equation}
where, $u = U - U_c$, $t = T - T_c$, $d = D - D_c$ and $a_n$ are linear functions in $u$ and $t$, $a_n(u,t) = \beta_n u + \gamma_n t$. The expansion to fourth order in $d$ is the minimal
model for a system with a cusp singularity.\cite{Golubitsky:1978aa} The free parameters of the fit are, $U_c$, $T_c$, $D_c$, $\beta_n$ and $\gamma_n$.

For fixed $u$ and $t$ the extremal points of $\mathcal{L}$,
\begin{equation}
  \partial_d \mathcal{L} = 	a_0 + a_1 d + a_2 d^2 + d^3 =0
  \label{eq:dLandau}
  \, ,
\end{equation}
are the Landau analogue of the DMFT fixpoints spanning a continuous surface $S_{\mathcal{L}} = \{(u,t,d):\,\partial_d \mathcal{L}(u,t,d) = 0\}$, in $(u,t,d)$ phase space. The hysteresis boundaries on $S_{\mathcal{L}}$ satisfy, $\partial_d^2 \mathcal{L} = 0$,  and in addition the critical point is characterized by, $\partial_d^3 \mathcal{L} = 0$, $\partial_d^4 \mathcal{L} > 0$.

To obtain a fit to the DMFT-ED data, on the $\pm1\,$meV scale around the critical point, a second order temperature term was added to $a_0(u,t)$,
\begin{align}
  a_0(u,t) = \beta_0 u + \gamma_0 t + \gamma_0^{(2)} t^2
  \, .
\end{align}
Note that this extra term does \emph{not} change critical behavior of the Landau functional, but changes the behavior of its ``unstable'' solution significantly.

We asses the ability of the minimal cusp singularity Landau functional to describe the DMFT fixpoints in the vicinity of the critical point by comparing, the $(U,T)$ phase diagram, the isotherms of $D(U)$ and $\epsilon(U)$ and finally the critical behavior along the first order transition line.

\begin{figure}
  \includegraphics{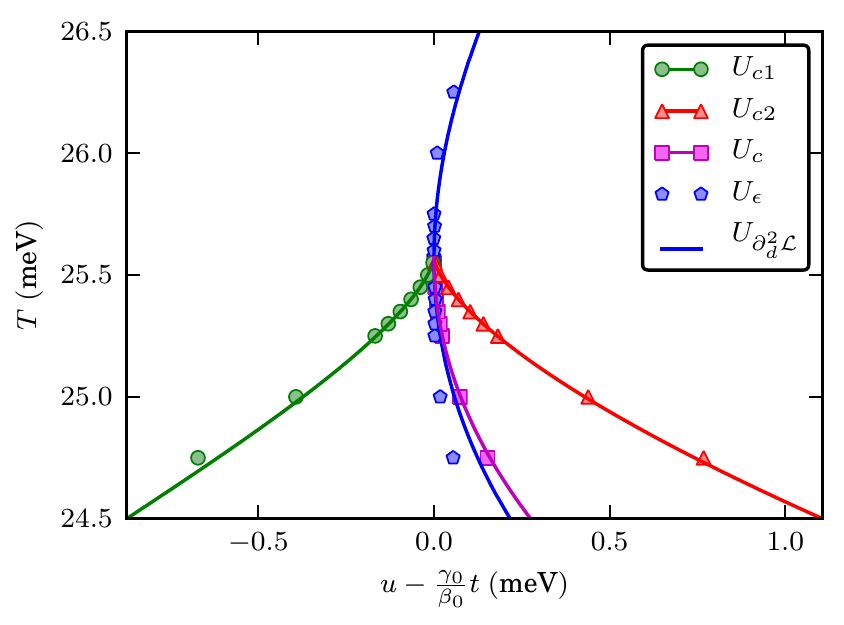}
  \caption{\label{fig:CritPointZoom} (Color online)
    Landau fit (solid lines) and DMFT-ED data (markers) in the $(u-\gamma_0 / \beta_0 t,T)$ plane close to the critical point.}
\end{figure}

When studying the phase diagram in the $(U,T)$ plane it is useful to compensate for the linear slope of the hysteresis region at the critical point. From the Landau functional the slope is obtained as the dotted line, $\beta_0 u + \gamma_0 t = 0$, shown in Fig.\ \ref{fig:PhaseDiag}. 
By the transformation, $u \rightarrow u - \frac{\gamma_0}{\beta_0}t$, this line becomes vertical and the corresponding transformed phase diagram is shown in Fig.\ \ref{fig:CritPointZoom}. The phase diagram of the DMFT-ED data and the Landau functional converge when
approaching the critical point.

\begin{figure}
  \includegraphics{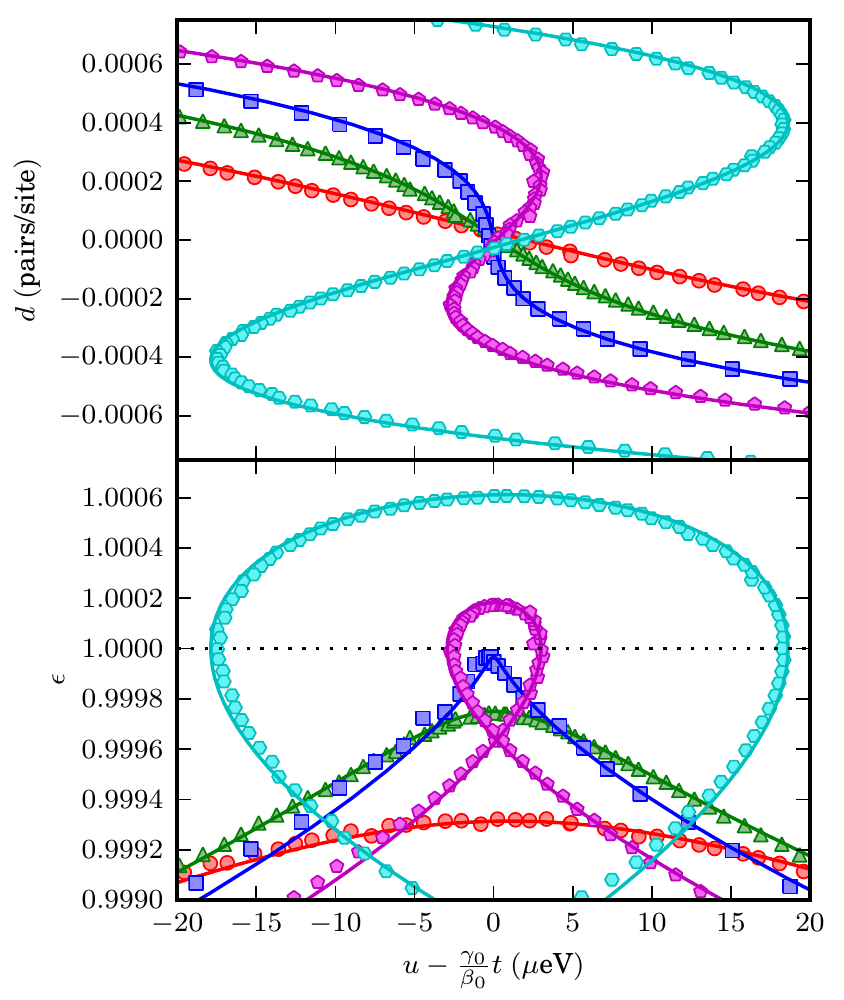}
  \caption{\label{fig:CritPoint} (Color online)
    Landau fit (solid lines) and DMFT-ED data (markers) in the vicinity of the critical point for the shifted double occupancy, $d=D-D_c$ (upper panel), and maximum eigenvalue $\epsilon$ of
$J_F(\Sigma^*)$ (markers) and the scaled and shifted second order derivative of the Landau functional $\partial_d^2 \mathcal{L}/C + 1$ (solid lines) (lower panel), plotted against, $u = U - U_c$, on isotherms with, $T= 25.0\,$meV + (0.650, 0.600, 0.575, 0.550, 0.500)$\,$meV, for \textcolor{red}{circles}, \textcolor{OliveGreen}{triangles}, \textcolor{blue}{squares}, \textcolor{magenta}{pentagons} and
\textcolor{cyan}{hexagons} respectively.}
\end{figure}

The Landau functional isotherms of $D(U)$ agree remarkably well with the calculated DMFT-ED data as shown in Fig.\ \ref{fig:CritPoint}. An important additional fact is that the largest eigenmode $\epsilon$ of the Jacobian $J_F$ of the DMFT fixpoint function $F_{U\beta}$ and the second derivative of the Landau functional $\partial_d^2 \mathcal{L}$ are related through,
\begin{align}
  \partial^2_d \mathcal{L} = C (\epsilon-1)
  \, ,
  \label{eq:EpsL}
\end{align}
for, $(u,t,d)\in S_{\mathcal{L}}$, where $C$ is a constant factor. (See the lower panel of Fig.\ \ref{fig:CritPoint}.) As only one eigenmode of the DMFT Jacobian $J_F(\Sigma^*)$ becomes critical in the hysteresis region, this mode alone governs the critical behavior of $D$, while all other eigenmodes gives the universal structure of the phase diagram around the critical point. The observed proportionality of Eq.\ (\ref{eq:EpsL}) confirms that our Landau functional, parametrized by the single parameter $d$, is able to describe the final ``effective'' critical behavior.

From Eq.\ (\ref{eq:EpsL}) we can construct $U_{\partial^2_d   \mathcal{L}}(T)$ analogously to $U_\epsilon(T)$ as the coupling maximizing $\partial^2_d \mathcal{L}$ on an isotherm with temperature $T$. Comparing $U_{\partial^2_d \mathcal{L}}(T)$ to $U_\epsilon(T)$, see Fig.\ \ref{fig:CritPointZoom}, qualitative agreement is achieved below $T_c$ and quantitative agreement above $T_c$.

\begin{figure}
  \includegraphics{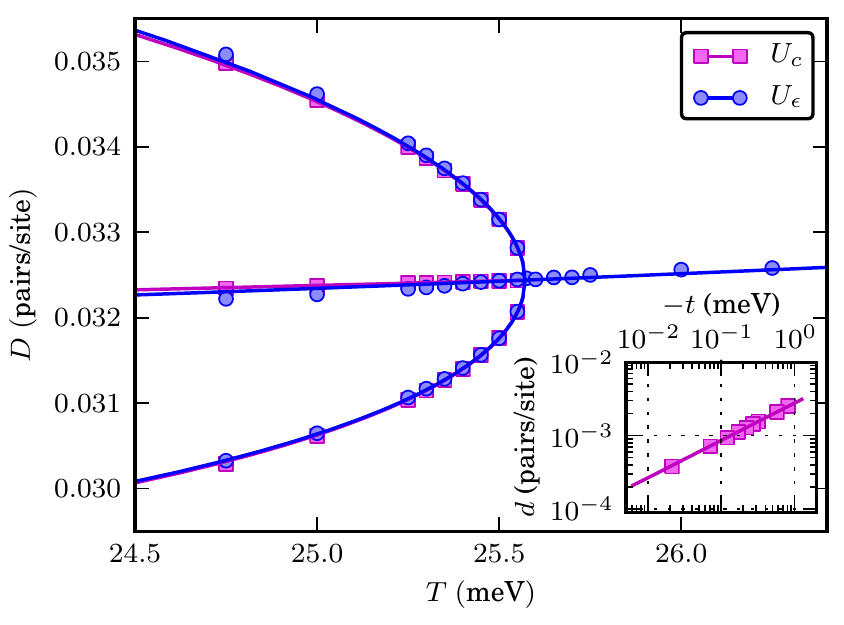}
  \caption{\label{fig:DvsT} (Color online)
    Double occupancy $D$ as a function of temperature $T$ on the $(U_c(T),T,D)$ and $(U_\epsilon(T),T,D)$ surfaces from DMFT-ED (markers) and the Landau fit (solid lines). (Inset) Logarithmic plot of the metallic branch of $(U_c(T),T,D)$ showing the $d\sim t^{\frac{1}{2}}$ critical exponent.}
\end{figure}

As a final test of the Landau fit we study the critical behavior of the fixpoints on the surfaces defined by the first order transition line $U_c(T)$ and the critical eigenmode maximum line $U_\epsilon(T)$. On these surfaces, $(U_c(T), T, D)$ and $(U_\epsilon(T), T, D)$, the emergence of the pitchfork bifurcation at $T_c$ is clear, and the Landau model and DMFT-ED data are again in good agreement. (See Fig.\ \ref{fig:DvsT}.)

We have shown that the Landau functional $\mathcal{L}$ quantitatively reproduces the critical properties of the DMFT fixpoint surface in the $(U,T,D)$ phase space and critical exponents can now be derived directly from $\mathcal{L}$. For, $U=U_c$, the critical behavior of the double occupancy with respect to $T$ is given by Eq.\ (\ref{eq:dLandau}) as,
\begin{equation}
  D - D_c \sim |T - T_c|^{\frac{1}{3}}
  \label{eq:DTCritical}
  \, ,
\end{equation}
and equivalently for $T=T_c$ the double occupancy as a function of $U$ has the same critical form,
\begin{equation}
  D - D_c \sim |U - U_c|^{\frac{1}{3}}
  \, .
\end{equation}

Along the first order transition line we have a very different critical behavior. Approaching criticality this line coincides with the first order term in the Landau functional being zero, $a_0(u,t) = \beta_0u + \gamma_0 t + \gamma_0^{(2)} t^2 = 0$. Sufficiently close to the
critical point only the linear order contribute giving, $u = -\frac{\gamma_0}{\beta_0}t$. With this constraint on $u$ we regain ``Ising'' scaling exponents and the temperature critical exponent changes to $1/2$ ie.,
\begin{align}
  D - D_c \sim |T - T_c|^{\frac{1}{2}}
  \, , \quad \textrm{iff } 
  U = U_c - \frac{\gamma_0}{\beta_0} (T  - T_c)
  \, ,
  \label{eq:DTCriticalTransline}
\end{align}
which is also confirmed by the DMFT-ED data and Landau fit in the inset of Fig.\ \ref{fig:DvsT}.

\begin{table}
\caption{\label{tab:CriticalPoint}
Comparison of $(U_c,T_c,D_c)$ for the second order critical end point.}
\begin{ruledtabular}
\begin{tabular}{lD{.}{.}{13}D{.}{.}{15}}
& \multicolumn{1}{c}{$U_c$ (eV)} & \multicolumn{1}{c}{$T_c$ (meV)}\\
\hline
This work, ED & 2.3398 \pm 0.0030\footnotemark[1] & 25.5625 \pm 0.0125\footnotemark[1] \\
HF-QMC\cite{Blumer:2002aa}& 2.3325 \pm 0.015 & 27.5\ \ \ \ \, \pm 0.2 \\
HF-QMC\cite{Rozenberg:1999aa}  & 2.38\ \ \ \pm 0.02 & 25.0\ \ \ \ \,  \pm 3.0 \\
ED\cite{Tong:2001aa} & 2.34 &  25  \\
NRG\cite{Bulla:2001yq}& - &  40  \\
\hline
This work, IPT & 2.46073 \pm 0.00050 & 46.9048 \pm 0.0550 \\
IPT\cite{Kotliar:2000aa} & 2.46315 & 46.895 \\
IPT\cite{Rozenberg:1994aa} & 2.51 & 44.0
\end{tabular}
\end{ruledtabular}
\footnotetext[1]{Errors given with respect to ED using $N_f=6$, not including
  finite size effects.}
\end{table}

The Landau fit also gives a very precise location of the critical point, in the approximation of DMFT-ED with, $N_f=6$, we obtain $D_c = 0.03244\pm 0.0001 \, \mbox{pairs/site}$, for $U_c$ and $T_c$ see Table \ref{tab:CriticalPoint}. The error in $T_c$ is estimated by the temperature difference between the isotherms closest to $T_c$ and the error in $U_c$ and $D_c$ are estimated from the isotherms in Fig.\ \ref{fig:CritPoint}. The values are compatible with previous reports on the critical point, see Table \ref{tab:CriticalPoint}. 
Even though the \emph{precision} is very high, the errors are estimated within the approximation of ED with a SIAM size of, $N_f = 6$, using the weight function of Eq.\ \ref{eq:chi2}. The \emph{real accuracy} is lower due to the finite size of the SIAM and the particular choice of weight function.

\begin{figure}
  \includegraphics{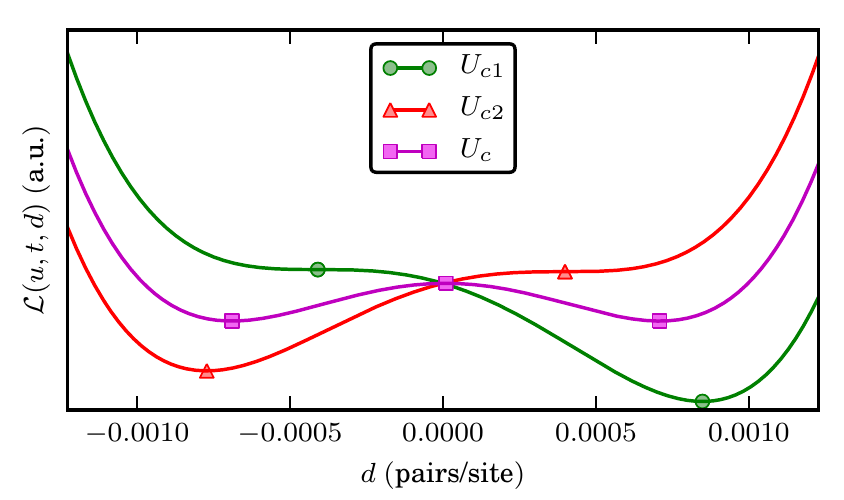}
 \caption{\label{fig:LandauEnergy} (Color online)
    The Landau function $\mathcal{L}$ as a function of the double occupancy, $d=D-D_c$, at $T = 25.5\,$meV, slightly below $T_c$, for $U$ fixed at $U_{c1}$, $U_{c2}$ and $U_c$. Markers indicate stationary points of $\mathcal{L}$.}
\end{figure}

The fixpoint surface $S_{\mathcal{L}}$ can be intuitively understood in terms of stationary points of the Landau functional $\mathcal{L}$ as a function of $d$. In Fig.\ \ref{fig:LandauEnergy}, $\mathcal{L}$ is shown for fixed $T$ and $U$ on the hysteresis boundaries and at the first order transition.
For $U = U_c$ the fixpoints correspond to two equal local minima and one unstable maxima. As the local minimas have the same value of $\mathcal{L}$ the system is unstable and can undergo a first order phase transition.
At the hysteresis boundaries $U = U_{c1}, U_{c2}$ a saddle-node bifurcation occur through the appearance of an inflection point, with two coinciding stationary points that separate when moving in to the hysteresis region. The local maxima corresponding to the unstable solution is always the stationary point with the highest free energy and can never be the thermodynamic ground-state.

\begin{figure}
  \includegraphics{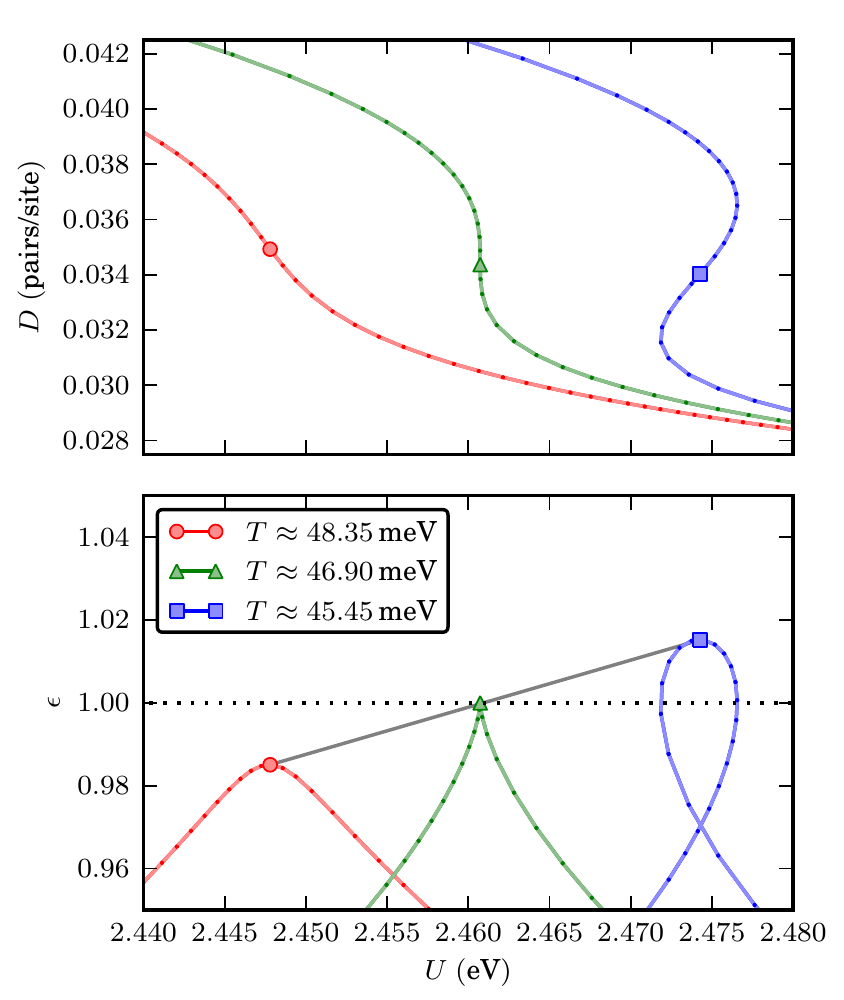}
 \caption{\label{fig:IPT_DEvsU} (Color online)
DMFT-IPT results for the double occupancy (upper panel) and maximum eigenvalue $\epsilon$ of $J_F(\Sigma^*)$ (lower panel) plotted against $U$ on isotherms above, close to and below the critical point (\textcolor{red}{circles}, \textcolor{OliveGreen}{triangles} and \textcolor{blue}{squares} respectively).}
\end{figure}

To demonstrate generality of the phase space extension and the critical properties of the MIT the same calculations have also been performed using the impurity solver IPT. In the IPT calculations the Matsubara formalism is implemented in terms of periodized Green's functions. 
In the ED calculations the DMFT problem $F_{U\beta}$ was formulated as a fixpoint problem in terms of the parametrized bath Green's function $G_0(i\omega_n)$, $F_{U\beta}(G_0) = F_{U\beta}(\mbf{x})$, while in the IPT calculations the fixpoint problem was formulated in terms of the self-energy $\Sigma(i\omega_n)$, $F_{U\beta}(\Sigma)$. The Jacobian was evaluated numerically and the maximal eigenvalue $\epsilon$ determined.
Applying the phase space extension to IPT we get exactly the same critical behavior as before, see Fig.\ \ref{fig:IPT_DEvsU}, although the location of the critical point is shifted as previously reported,\cite{Kotliar:2000aa, Rozenberg:1994aa} see Tab.\ \ref{tab:CriticalPoint}.

\section{\label{sec:Discussion}Discussion}

Using the phase space extension and the Jacobian $J_F(\Sigma^*)$ of the DMFT fixpoint function we have mapped out the metal insulator phase digram, the different DMFT fixpoints and their bifurcations. This extends previous calculations\cite{Kotliar:1999aa,Kotliar:2002aa} by describing the unstable solution, which allows us to compute a continuous two dimensional fixpoint surface in the three dimensional $(U,T,D)$ phase space.

The critical point has been classified as a cusp catastrophe with a pitchfork bifurcation on the first order transition line.  The hysteresis boundaries have been shown to be saddle-node bifurcations of two merging fixpoints, one stable and one unstable respectively.
Using the explicit calculation of $J_F(\Sigma^*)$ we have confirmed the prediction that a single critical eigenmode governs the MIT in DMFT\cite{Kotliar:2000aa} and shown that this mode becomes critical not only at the critical point but also on the hysteresis boundaries. 
The calculated fixpoint surface was then used to fit a cusp catastrophe minimal Landau functional expansion in the vicinity of the critical point. 

The excellent agreement between the mean field Landau model and the DMFT-ED data shows that the MIT critical point in DMFT do have mean-field critical behavior, confirming previous reports.\cite{Kotliar:1999aa,Kotliar:2000aa,Kotliar:2002aa}
From the Landau fit the critical temperature and coupling, $T_c$ and $U_c$, was determined with high precision but with an accuracy limited by finite size effects of the ED impurity solver.

The study of the DMFT fixpoint surface using the phase space extension and the Jacobian has shown all the general features of the MIT. In principle it can be combined with numerical exact impurity solvers like Continuous Time Quantum MonteCarlo (CT-QMC) to remove convergence issues caused by fixpoint-bifurcations. This combination has the potential to further improve the accuracy in the location of the critical point. 

Our ED results for the critical point agrees with previous ED calculations\cite{Tong:2001aa} using the same number of bath sites, see Tab.\ \ref{tab:CriticalPoint}. Also the results of previous Hirsch-Fye QMC calculations\cite{Blumer:2002aa,Rozenberg:1999aa} are compatible with our ED results. 
The exact position of the critical point from IPT does not converge to the one of the other impurity solvers. This cannot to be expected since the expression for the self-energy is truncated at second-order in $U$. However our discretization procedure yields results consistent with other IPT calculations.\cite{Kotliar:2000aa, Rozenberg:1994aa}

Our results, showing that the fourth order Landau functional describes the critical properties of the DMFT solution when using ED, confirms the findings of Kotliar {\it et.\ al.}\cite{Kotliar:2000aa} who showed, using Hirsch-Fye QMC and IPT, that the critical properties are not impurity solver dependent.

Using the Landau functional we obtain the same universal $1/3$ critical exponents, in Eq.\ (\ref{eq:DTCritical}), as initially reported for DMFT\cite{Kotliar:2000aa} and later also found experimentally\cite{Limelette:2003kx} in Cr doped V$_2$O$_3$ in the $U$ and pressure dependence respectively.
We also present DMFT results on the $1/2$ critical exponent, Eq.\ (\ref{eq:DTCriticalTransline}), along the first order transition line. This has been found experimentally in the temperature dependence, $D - D_c \sim |T - T_c|^{\frac{1}{2}}$, of Cr doped V$_2$O$_3$, where the coupling to temperature in the first order term of the Landau functional vanishes, $\gamma_0/\beta_0 \approx 0$.

Regarding the minimal cusp singularity Landau model, Eq.\ (\ref{eq:Landau}), it is noteworthy that the linear expansion of the $a_n(u,t)$ coefficients captures the critical behavior of the metallic and insulating solution of the DMFT data.
However, an additional second order temperature term $\gamma_0^{(2)}t^2$ must be added to correctly describe the unstable solution. Without this term, the Landau line $U_\epsilon(T)$ in Fig.\ \ref{fig:CritPointZoom} becomes a straight line, all isotherms of the Landau functional cross the same point $(u-\gamma_0/\beta_0t, d) = (0,0)$ (Fig.\ \ref{fig:CritPoint}) and the unstable central branch of the pitchfork bifurcation in Fig.\ \ref{fig:DvsT} becomes horizontal. 

\section{\label{sec:Conclusion}Conclusion}

In this paper we have presented the phase space extension algorithm for solving the single band DMFT fixpoint problem that is free from the numerical problems experienced by other methods in the hysteresis region of the MIT.

We have also performed an explicit calculation of the Jacobian of the DMFT fixpoint function and its corresponding eigenmodes. Using the critical eigenmode of the Jacobian we explained the critical slowing down and inability to find the thermodynamically unstable solution using forward recursion. Moreover the instability of both forward recursion and Newton algorithms on the hysteresis boundaries has been explained in terms of one eigenmode going critical.

The critical properties of the second order critical point of the MIT has been shown to be representable by a mean field Landau functional, giving the general properties of the DMFT fixpoint surface in $(U,T,D)$ phase space in terms of a cusp singularity with a pitchfork bifurcation and saddle-node bifurcations on the hysteresis boundaries. Experimentally the pressure driven MIT in Cr-doped $V_2O_3$ show the same mean-field critical exponents for the second order critical point.\cite{Limelette:2003kx}

Finally we note that the phase space extension algorithm is general and can be combined with any impurity solver, and possibly also extended to study phase transitions in other systems such as the multi-band Hubbard model and the Hubbard-Holstein model.

\begin{acknowledgments}

Funding from the Mathematics - Physics Platform ($\mathcal{MP}^{\textsf{2}}$) at the University of Gothenburg and the Swedish Research Council (grant no.\ 2007-5397 and 2008-4242) is gratefully acknowledged.
The simulations were performed on resources provided by the Swedish National Infrastructure for Computing (SNIC) at Uppsala Multidisciplinary Center for Advanced Computational Science (UPPMAX) (project no.\ p2008033) and at Chalmers Centre for Computational Science and Engineering (C3SE) (project no.\ 001-10-37).
The authors would like to thank Dr.\ Ansgar Liebsch and Dr.\ Bernhard Mehlig for stimulating and valuable discussions.

\end{acknowledgments}

\bibliography{/Users/hugstr/Documents/Papers/DMFT_Biblography}

\end{document}